# Texture of ZnO thin films deposited through microwave irradiation

Piyush Jaiswal[1]*, Sanjaya Brahma[1], K. S. Suresh[2], S. A. Shivashankar[1], Satyam Suwas[2]

[1]Materials Research Centre, [2]Department of Materials Engineering, Indian Institute of Science, Bangalore – 560012, India

## Abstract

We report the crystallographic texture of ZnO thin films comprising nanorods grown by a microwave irradiation-assisted method. Different substrates were used, namely Si, Ge, metal-coated Si, PMMA- coated Si and ITO-coated glass, to examine the respective development of crystallographic texture. The films were characterized by X–ray diffraction and scanning electron microscopy, while the texture analysis was done by X–ray pole figure analysis using the Schultz reflection method. The ZnO film deposited on Si and Ge showed a mixed texture. On all the other substrates, c–axis oriented ZnO films were obtained.

*Corresponding author: Tel - +91-80-22932782, Fax – +91-80-23607316

*E. mail Address*: piyush@mrc.iisc.ernet.in

## 1. Introduction

ZnO is known to exhibit a diverse and rich variety of nanostructures and has attracted much attention in recent years because of its unique electrical, optical, and catalytic properties. The electrical [1,2] and optical properties [3] of the films depend on their microstructure, crystallographic texture and surface morphology, as well as on the presence of defects. For example ZnO thin film grown on a C- plane sapphire has better structural and optical quality than those grown on A – plane and R – plane sapphire [3]. The piezoelectric properties of ZnO thin film depend mainly on the crystallographic texture. The degree and type of texturing in thin films depend on various factors, such as substrate temperature, composition of the reaction mixture in case of liquid phase deposition, gas pressure and composition in case of gas phase deposition, and the growth rates of different crystallographic facets. The control of the crystallization and growth direction, which manifests itself to crystallographic texture, is an essential requirement for practical applications. For example, growth of (002)-oriented ZnO nanowires has practical applications such as in field emission [4], light emitting diodes, sensor arrays [5], and nanogenerators [6]. For a particular application, the film needs to be oriented in a specific growth direction, for example, ZnO thin film with (110) growth direction has been proposed for transparent conductive oxide applications [7], as well as for acousto-optical and piezoelectrical devices [8]. It is, therefore, almost mandatory to know the details of texture formation in ZnO thin films as a function of processing variables. A variety of thin film deposition techniques have been developed for the growth of ZnO thin films, for example, atomic layer deposition (ALD) [9], molecular beam epitaxy (MBE) [10], pulsed laser deposition (PLD) [11], radio-frequency magnetron sputtering [12] and sol–gel method [13]. However, all these processes either require expensive equipment involving a vacuum system or call for a tedious synthesis procedure followed by post-synthesis heat treatment and, thus, are time consuming. In this regard, microwave irradiation-assisted deposition in the liquid medium is a more economical and easily applicable method for the preparation of such films. However, the formation of textures in so-obtained films is not known. Texture formation in this film, in general, is known to depend largely on the nature of the substrate in addition to other processing variables such as choice of a substrate and its orientation [3,14], laser power flux in a pulsed laser deposition [15], pre-coated seed layer of ZnO [16] etc. It is, therefore, worthwhile to study the texture of ZnO thin films deposited on different substrates by the microwave irradiation-assisted method [17]. This method being a rapid deposition process, *i.e.*, deposition takes place within a few minutes, is likely to give rise to different microstructure and texture in the resulting thin films from those in other deposition techniques. In the present work, zinc oxide thin films comprising nanorods have been deposited on semiconducting substrates such as Si(100) and Ge(100), on metal-coated Si, PMMA-coated Si(100), and ITO-coated glass. The textures formed have been analyzed by through pole figures method obtained by the X-ray diffraction technique. The textures have been quantified by orientation distribution function method.

## 2. Experimental details

One gram of a high-purity zinc acetylacetonate ($C_{10}H_{14}O_4$. Zn) was taken in a 250 ml round-bottomed flask and dissolved in ~40 ml of ethanol (99.9%, HPLC-grade) and stirred for fifteen minutes. A dilute solution of a surfactant *e.g.,* polyvinylpyrrolidone (abbreviated as PVP, mol.wt. 360,000) in double-distilled water was prepared separately (~0.3 g PVP in 40 ml water) and added to the above “precursor solution”, followed by another fifteen minutes of stirring. In this consolidated solution, taken in a round-bottomed flask, was suspended a substrate, such as Si(100), measuring up to 15 mm x 15 mm. The round-bottomed flask was then placed in a domestic-type microwave oven (operating at 2.45 GHz, with power variable from 160 – 800 W) and the resultant mixture subjected to microwave irradiation at 800 W for 5 minutes. Microwave irradiation of the solution, was found to result in a colloidal suspension in the solution and a coating on the substrate. The substrate was removed carefully from the solution and washed with distilled water and acetone. Upon drying, a dull white coating was visible on the substrate. The above procedure was subsequently repeated separately, to obtain similar coatings on Ge(100) and ITO-coated glass substrates. ZnO coatings were also obtained by a similar procedure on Cr-coated Si and PMMA-coated Si, but employing CTAB as surfactant (cetyl trimethylammonium bromide). These coatings are found to adhere well to the respective substrate, as revealed by the peel-tape test [18].

The resulting coated samples were characterized by powder X-ray diffraction (XRD) in a Bruker D8 Advance instrument using Cu-$K_\alpha$ radiation. XRD patterns were recorded from 20° to 90° (2θ) with a scanning step of one degree per minute. The morphology of the coatings was examined by field-emission scanning electron microscopy (FE-SEM SIRION XL-40 operating at 40 kV).

Thin film X-ray texture analysis was carried out on the films obtained under different deposition conditions to examine the presence of crystallographic texture in the deposited films using Bruker D8 Discover texture goniometer with Cu-$K_\alpha$ radiation.

Crystallographic texture of the films was determined by measuring (100), (002) and (101) pole figures using Schultz's reflection method up to 70 ° tilt angle. The data so obtained were employed for the calculation of the complete Orientation Distribution Function (ODF) using the arbitrary defined cell method. The ODFs were calculated using Labotex V3.0 software. From the ODFs, the relevant pole figures were recalculated with an axial symmetry. This was done to avoid the possible omission of poles in the outer regions of the pole figures. For the calculation of the three-dimensional ODFs, orthorhombic sample symmetry was applied. Although pole figures and ODFs represent the texture quite sufficiently, it is sometimes required to express the textures in terms of geometrical axis of the sample. This is done by expressing the texture in terms of inverse pole figures (IPF). The volume fractions of relevant texture fibers, namely, <100>, <002>, and <101> were also calculated using the Labotex software using an orientation spread of 5°.

## 3. Results and Discussion

### *3.1. Structural and microstructural characterization*

The crystallinity of the films deposited on variety of substrates using the procedure described above was established by X-ray diffraction. The XRD pattern of the ZnO film deposited on Si(100) is shown in Fig. 1, which indicates clearly that the film was crystalline in nature. The pattern shows high intensity peaks at different Bragg angles, which indicates that the film is polycrystalline. The pattern can be indexed to that of the hexagonal würtzite phase of ZnO. The absence of any peaks other than those due to ZnO and Si(100) indicates that the chemical reactions which take place under microwave irradiation lead to no other “impurity” phases in the film deposited, to within the detection limits of powder XRD. (The remnants of the surfactant will have been “burned off” by a brief heat treatment of the deposit in air at 500 ºC.). Similarly, polycrystalline films were obtained when Ge(100) is used as the substrate as shown in the XRD pattern Fig. 2(a). However, a (002)-oriented film was obtained when ITO-coated glass is used as the growth surface. Fig. 2(b) shows the XRD pattern of the ZnO film deposited on ITO-coated glass. The intensity of the (0002) peak is noticeably higher than that of all other peaks in the pattern, indicating that the film has preferred c-axis orientation, i.e., along (0002). There is a significant difference in the XRD pattern when Cr-coated Si or PMMA-coated Si are used as the growth surface, *i.e*, the ZnO film has a strong c–axis orientation (Fig. 2c,d).

The morphology of the as-prepared ZnO film grown on Si(100) is shown in the scanning electron micrographs of Fig. 3, of which Fig. 3(a) shows coating of ZnO obtained at 800 W of power applied for 5 minutes. The coating comprises hexagonal nanorods with distinctively flat faces and nearly uniform diameters. The cross-sectional SEM (Fig. 3b) reveals that each nanorod, more than one µm long, is anchored to the Si substrate. Fig. 3(a) shows the presence of nanorods, whose growth direction is the c-axis [14]. Although their growth direction remains the same, the nanorods are inclined with respect to the substrate normal at somewhat random angles. Similar random inclination of nanorods with respect to the substrate has also been obtained in case of ZnO thin film deposited on Ge(100) substrate as seen in the SEM image (Fig. 4(a)). Fig. 4(b) shows the SEM image of ZnO nanorods grown on ITO-coated glass. Here, it is seen that the film has a preferred c-axis orientation *i.e.,* many nanorods are grown nearly perpendicular to the substrate. However, in case of ZnO films grown on Cr-coated Si and PMMA-coated Si the film are found to be strongly oriented in the (0002) direction *i.e.*, most nanorods are grown perpendicular to the substrate, as shown by the top view of the ZnO nanorods (Fig. 4(c)), where the substrate is Cr-coated Si. Fig. 4(d) shows the top view of ZnO nanorods deposited on PMMA-coated Si(100) where only the tips of the nanorods are visible.

### *3.2 Characterization of crystallographic texture*

X-ray pole figures were measured to determine the crystallographic texture of the ZnO films deposited on different substrates. Fig. 5&6 show the recalculated (0002) and $(10\bar{1}0)$ pole figures, along with the inverse pole figures for ZnO thin films deposited on different substrates. From these two figures, it can be stated that the textures of the ZnO films deposited on ITO, PMMA-Si and Cr-coated Si substrates are qualitatively similar. The other two samples, namely the films on Si and Ge, are somewhat different. The recalculated PFs and IPFs (Fig. 5) for films deposited on on Si and Ge display a strong $(10\bar{1}0)$ fiber, along with a weak (0002) fiber. In addition to $(10\bar{1}0)$ and (0002) fibers, the film grown on Ge(100) indicates a strong $(21\bar{3}0)$ fiber also, which is clearly visible from the IPFs. The (0002) pole figures show that all the three samples are characterized by a strong basal texture with PMMA-coated Si substrate exhibiting the strongest basal pole, followed by Cr-coated Si and then ITO-coated glass (Fig. 6). The spread of basal fiber is greater in the films grown on Cr-coated Si and ITO-coated glass than in the film on the PMMA-coated Si substrate. Amongst the three, only the film deposited on PMMA-coated Si shows a small fraction of $(10\bar{1}0)$ fiber, while in the other two cases this fiber is not observed.

To have a better understanding of the texture evolution of ZnO thin films on different substrates, three-dimensional Orientation Distribution Functions (ODF) were plotted. Fig. 7 & 8 show these 3 ODF plots. The ODF was calculated with orthorhombic sample symmetry. As stated earlier, the effect of substrate on the growth texture of ZnO can be classified into two categories. For the film belonging to the first category (Fig. 7), certain components like $(10\bar{1}0) < 10\bar{1}0 >$ and $(10\bar{1}0) < 10\bar{1}0 >$ exhibit strong intensity. In the second category, where the films have exhibited strong basal texture, the maximum intensity corresponding to the $(0001) < 23\bar{5}0 >$ component was observed in the film deposited on ITO-coated glass. The samples deposited on PMMA-coated Si or Cr-coated Si display uniform intensity throughout the (0002) basal fiber.

The volume fractions of different hcp fibers are given in Fig. 9. These were calculated with a spread of 5° along $\Phi$ and $\varphi_2$. The ZnO film on PMMA-coated Si substrate shows the highest intensity of basal fibers among different samples, followed by the films on Cr-coated Si and the one deposited on ITO-coated glass. However, the spread of basal pole in the film deposited on Cr-coated Si is less than in the film on PMMA-coated Si. This is clearly visible from the pole figures (Fig. 6) as well as from the 3D ODF (Fig. 8). For the other two samples, the fraction of basal fiber is less than it is for the $(10\bar{1}0)$ and $(11\bar{2}0)$ fibers.

While the growth of strongly oriented ZnO nanorods takes place only under certain conditions, it is quite remarkable that, without any templating, strongly oriented growth takes place at all, especially considering that the growth process in the present experiments is rapid in comparison with many other techniques. It is to be noted that, in all cases, the nucleation density in the liquid medium on the substrate is very high. Thus, where strong orientation is observed, simultaneous formation of a very large number of similarly oriented nuclei takes place on the substrate, in the absence of any templating and, certainly, in the absence of any lattice matching. Perhaps this is driven by minimization of energy at the substrate-film interface [19] on smooth, disordered surfaces such as that of Si(100) coated by spun-on PMMA. However, an alternate, unknown mechanism is likely to be responsible for the strongly oriented growth on the smooth, polycrystalline, and electrically conducting surfaces of Cr-coated Si and ITO-coated glass. It is surmised that this mechanism is related to the complex interaction of the incident microwave radiation with these conducting surfaces.

## 4. Conclusion

The crystallographic texture of ZnO thin films deposited on different substrates through microwave-assisted chemical reaction in liquid medium has been determined. It is possible to tailor the texture in ZnO films by varying the nature of the substrate and the surfactant. A variety of basal textures ranging from weak to strong, as well as non-basal textures can be obtained.

**Acknowledgement**

The authors acknowledge the uses of the facility set up under Institute Nanoscience Initiative sponsored by DST-FIST program and X-ray facility. Sanjaya Brahma thanks the Council of Scientific and Industrial Research (CSIR) for the award of a research associateship.

**Figure captions**

Fig. 1. X- Ray diffraction pattern of ZnO deposited on Si(100).

Fig. 2. X- Ray diffraction pattern of ZnO deposited on (a) Ge(100), (b) ITO-coated glass, (c) Cr-coated Si(100), and (d) PMMA-coated Si(100).

Fig. 3. SEM image of ZnO nanorods deposited on Si(100): (a) top view (b) cross- section.

Fig. 4. SEM image (top view) of ZnO nanorods deposited on (a) Ge(100) (b) ITO- coated glass (c) Cr coated Si(100) and (d) PMMA-coated Si(100).

Fig. 5. X-ray pole figures data for ZnO coatings on (a) Si(100) (b) Ge(100) Deposition parameters were: *microwave irradiation time = 5 min, microwave power= 800 W, aqueous solution of surfactant (PVP, mol.wt. 360,000) and* $Zn(acac)_2$ *as precursor material dissolved in ethanol*.

Fig. 6. X-ray pole figures data for ZnO coatings on (a) ITO coated glass, (b) PMMA-coated Si(100) and (c) Cr-coated Si(100). Deposition parameters were: *microwave irradiation time = 5 min, microwave power= 800 W, aqueous solution of surfactant (CTAB) and* $Zn(acac)_2$ *as precursor material dissolved in ethanol.*

Fig. 7. 3D ODF for ZnO coatings on (a) Si(100), (b) Ge(100).

Fig. 8. 3D ODF for ZnO coatings on (a) ITO-coated glass, (b) PMMA-coated Si(100) and (c) Cr-coated Si(100).

Fig. 9. Volume fractions of different hcp fibers.

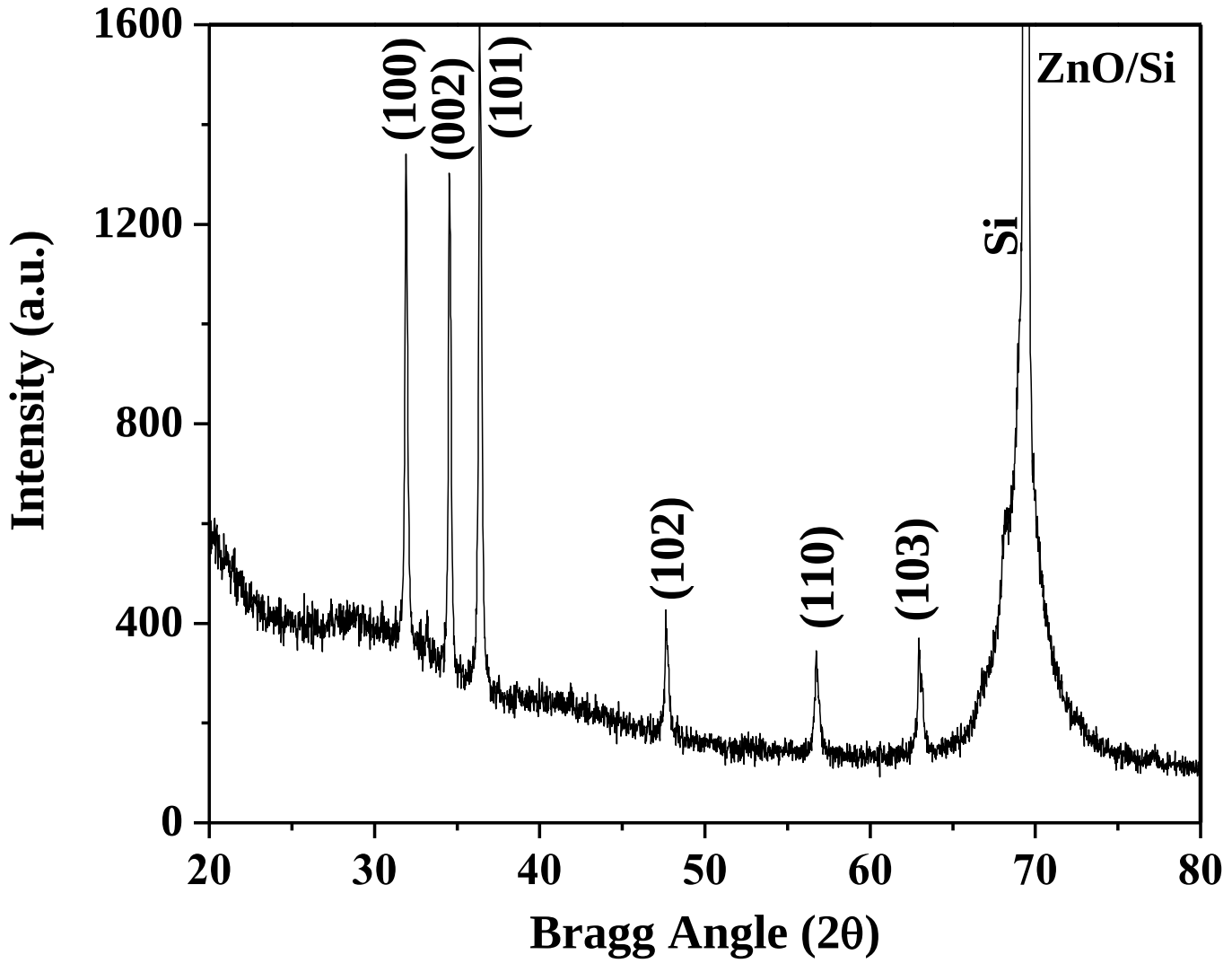

**Fig. 1.**

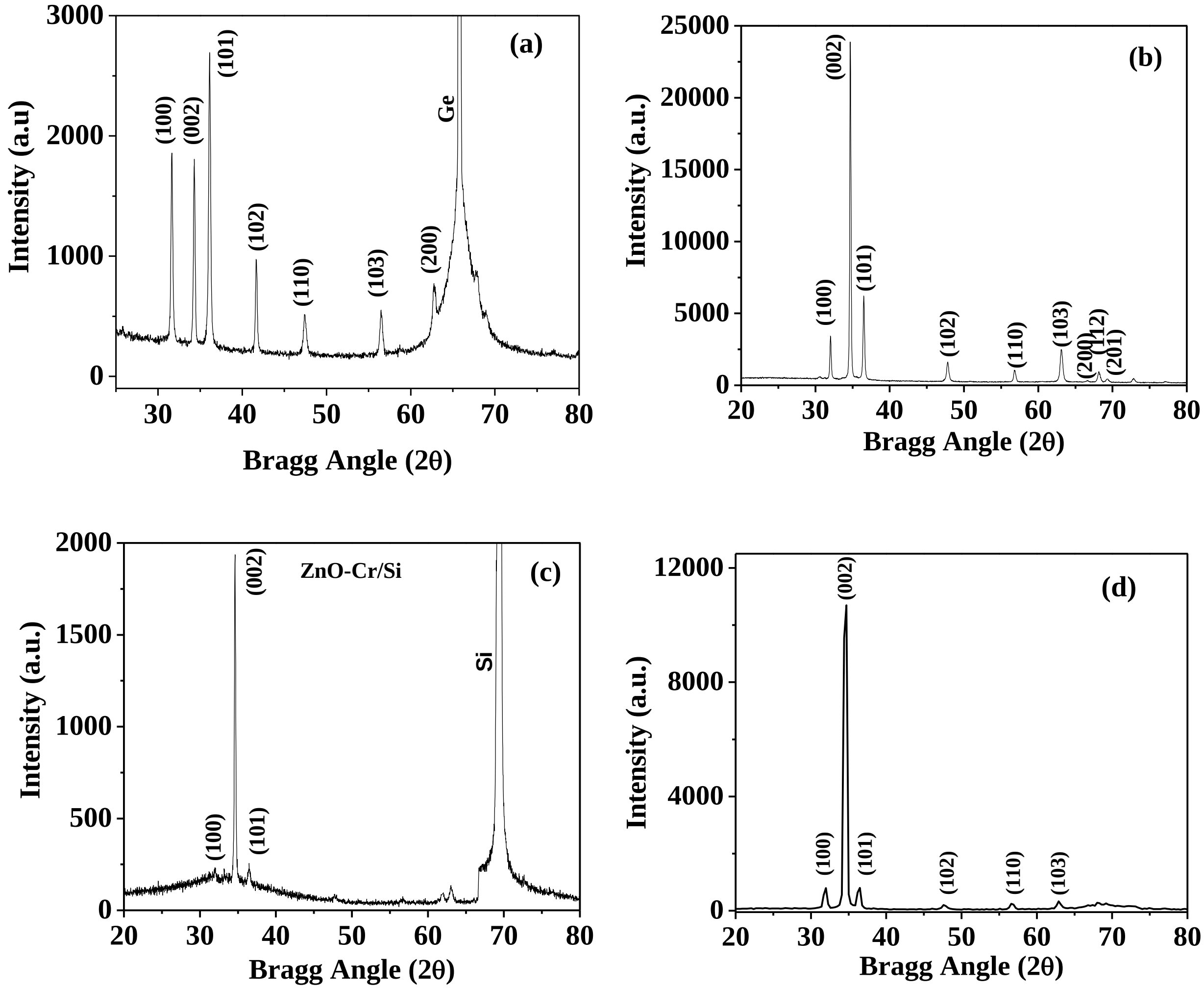

**Fig. 2.**

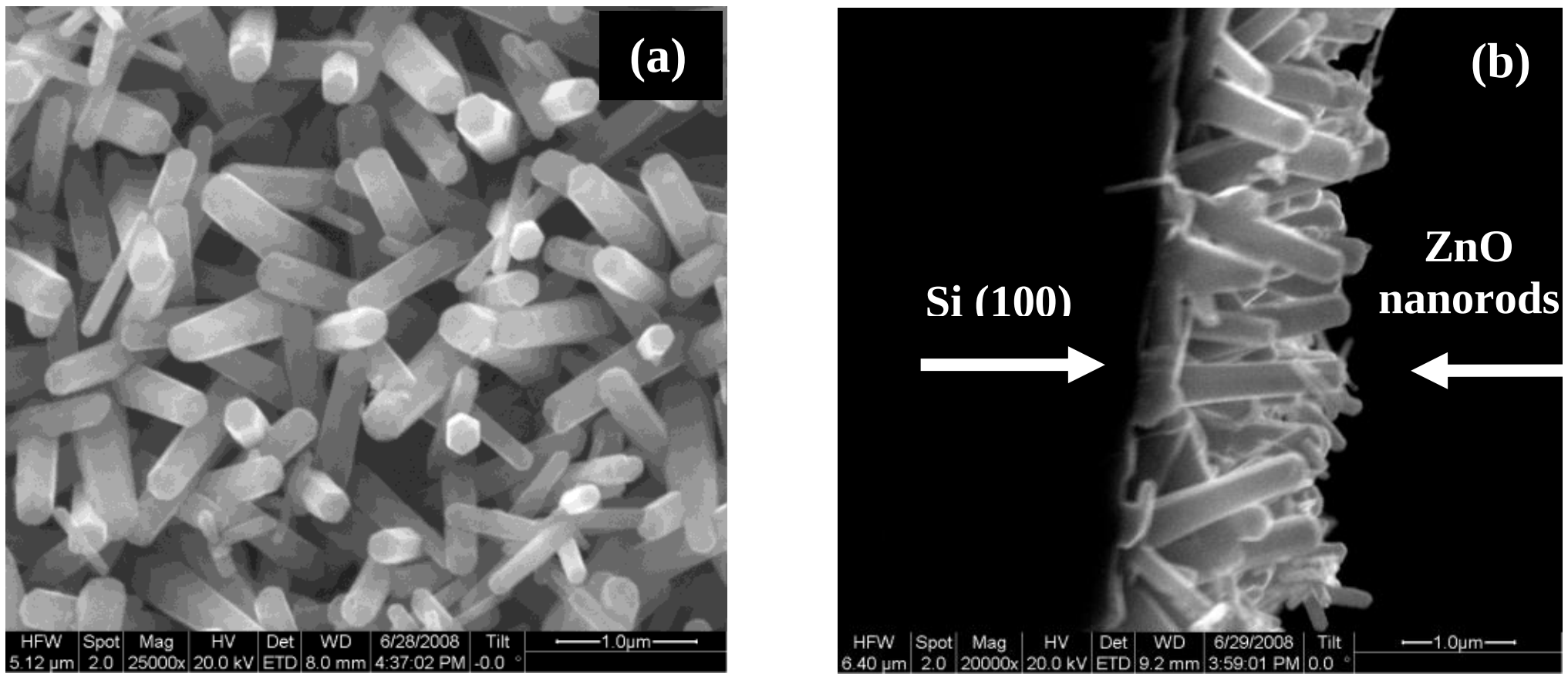

**Fig. 3.**

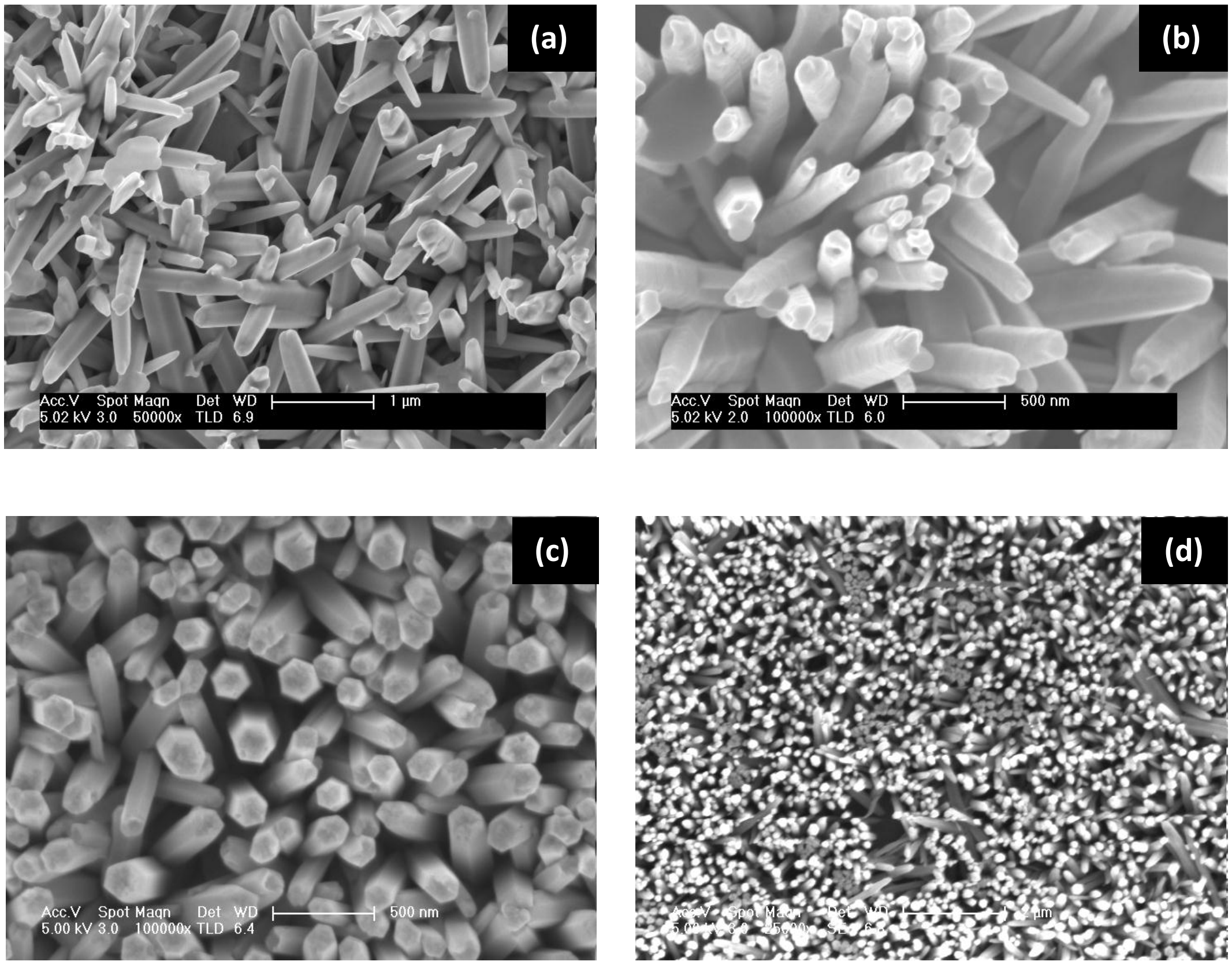
(a)
Acc.V Spot Magn Det WD
5.02 kV 3.0 50000x TLD 6.9
1 µm
(b)
Acc.V Spot Magn Det WD
5.02 kV 2.0 100000x TLD 6.0
500 nm
(c)
Acc.V Spot Magn Det WD
5.00 kV 3.0 100000x TLD 6.4
500 nm
(d)

**Fig. 4.**

| | ZnO-PVP-Si | ZnO-PVP-Ge | Level |
|---|---|---|---|
| IPF | 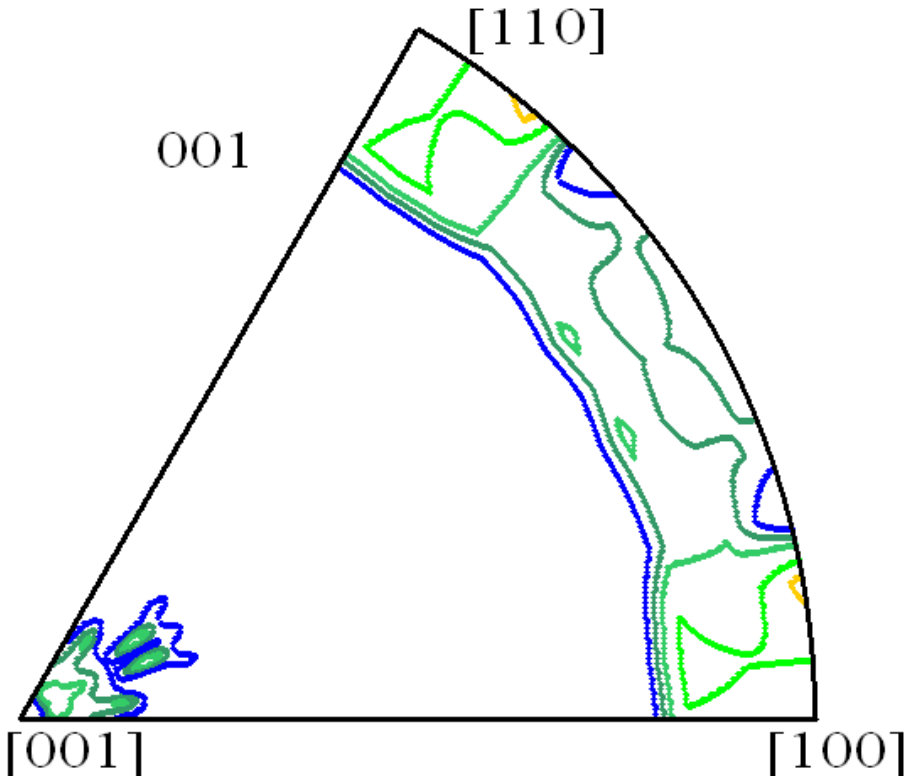  | 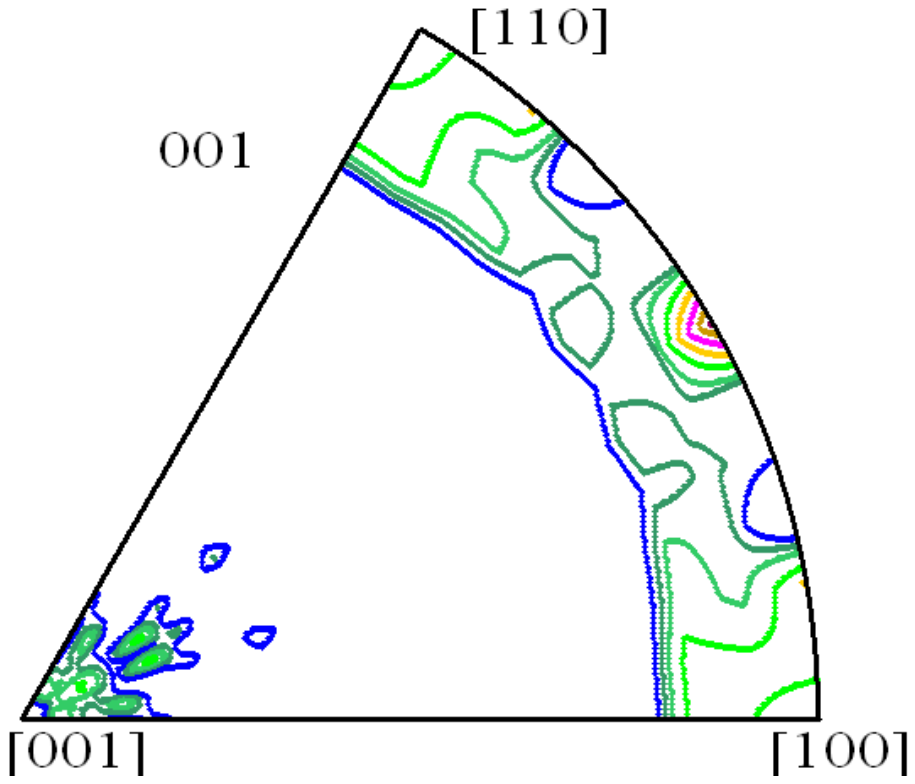  | 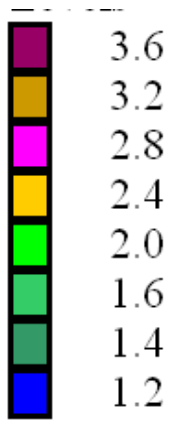  |
| Max | 2.6 | 3.7 | |
| PF<br>(100) | 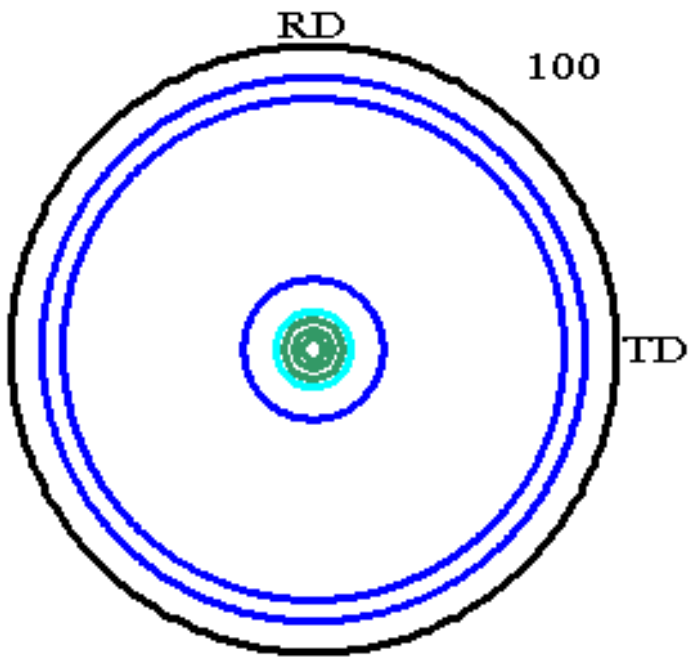  | 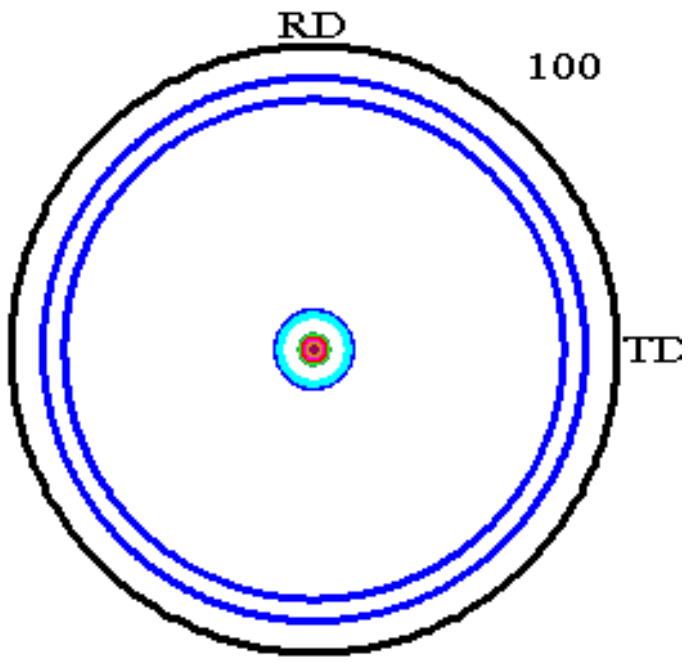  | |
| Max | 1.6 | 3.9 | |
| PF<br>(002) | 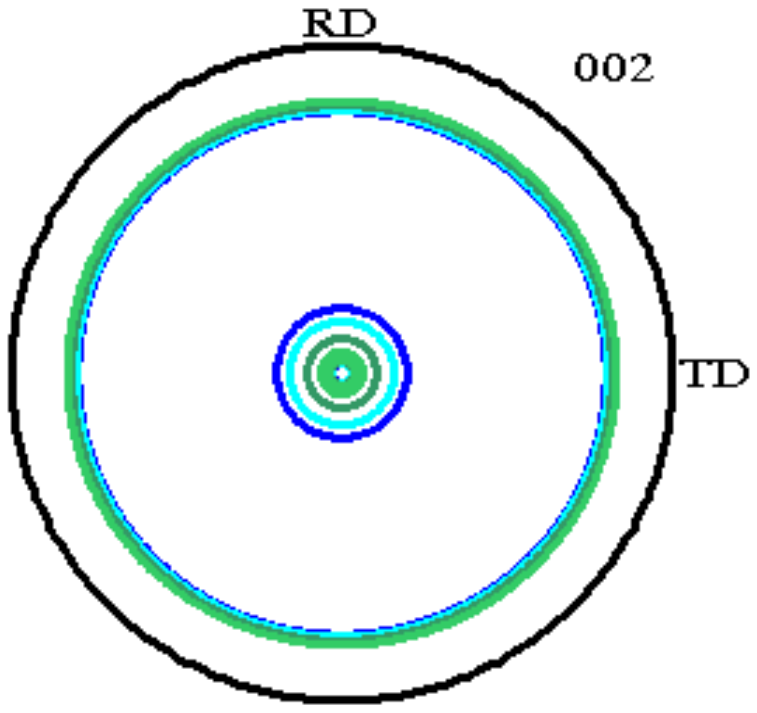  | 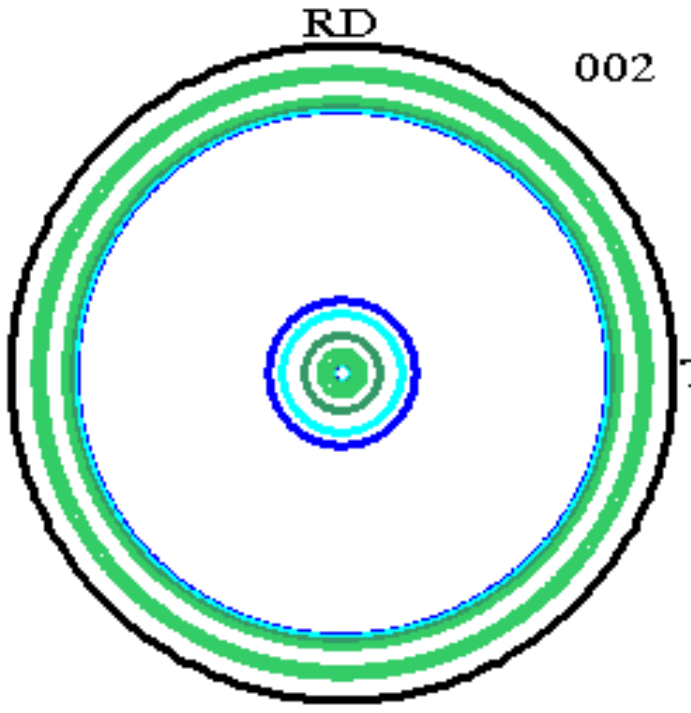  | 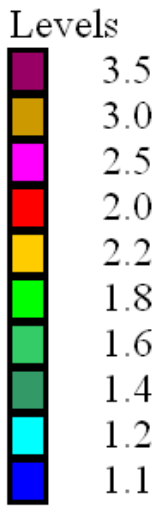  |
| Max | 1.7 | 1.7 | |

**Fig. 5.**

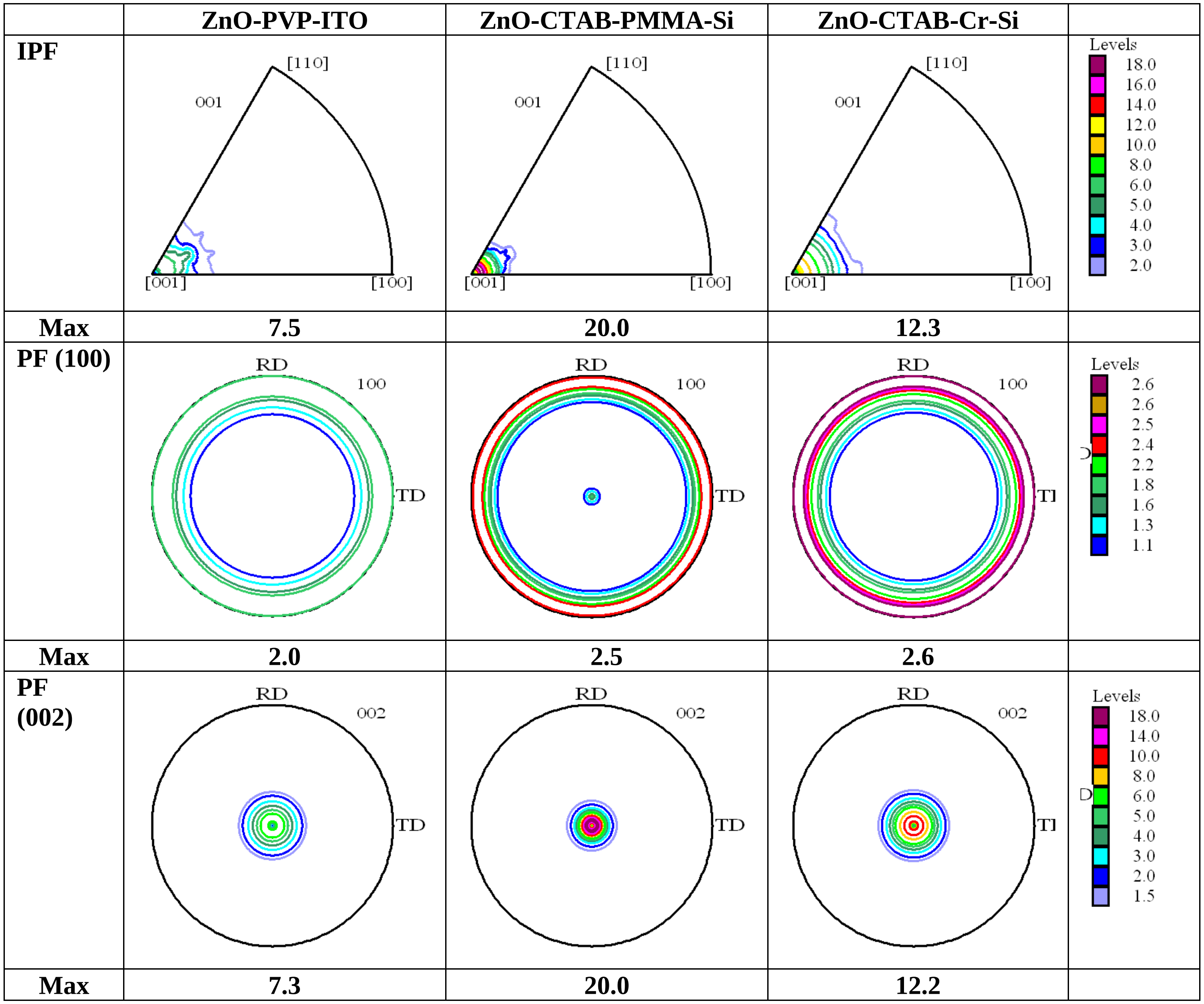

Fig. 6.

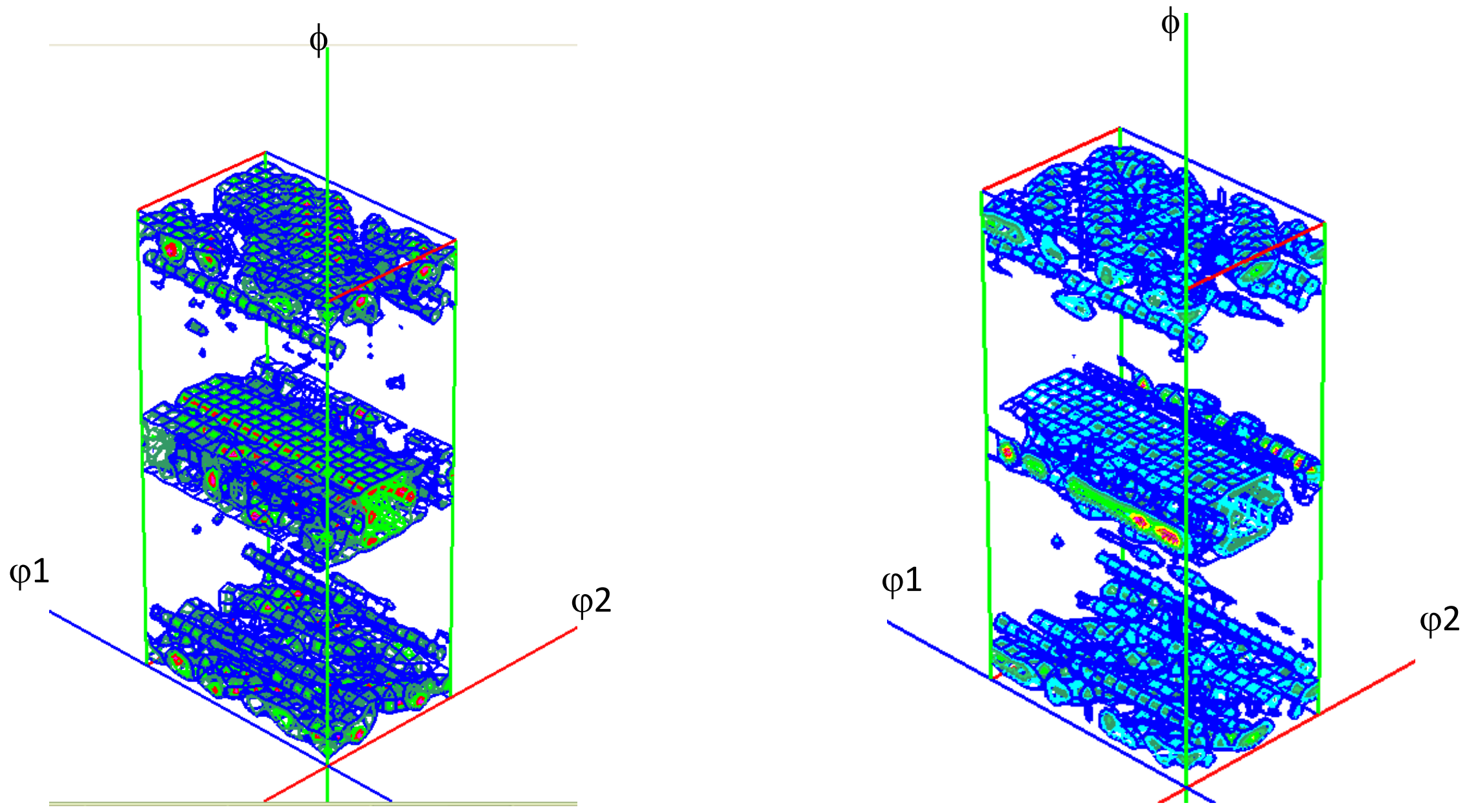

(a) (b)

**Fig. 7.**

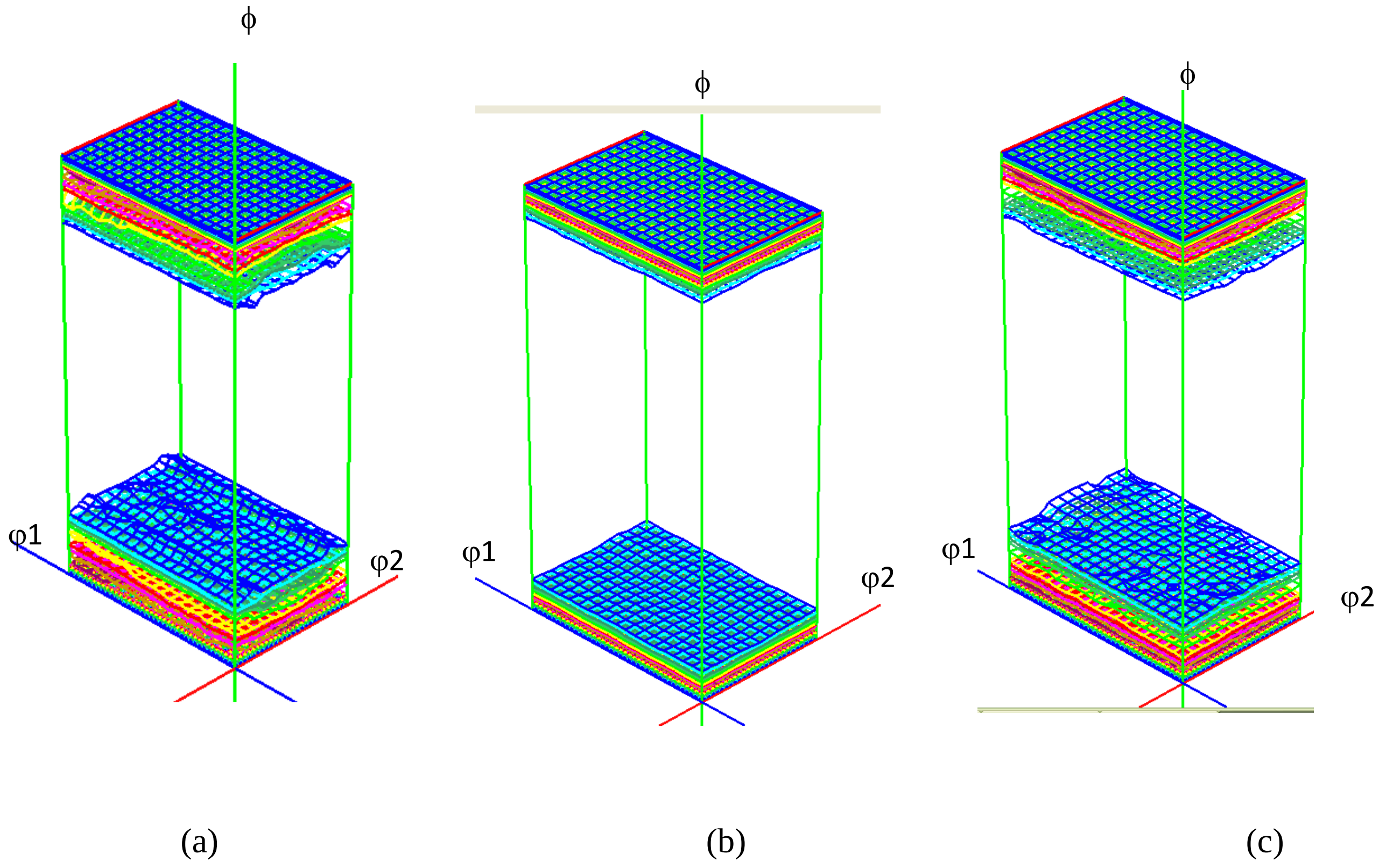

**Fig. 8.**

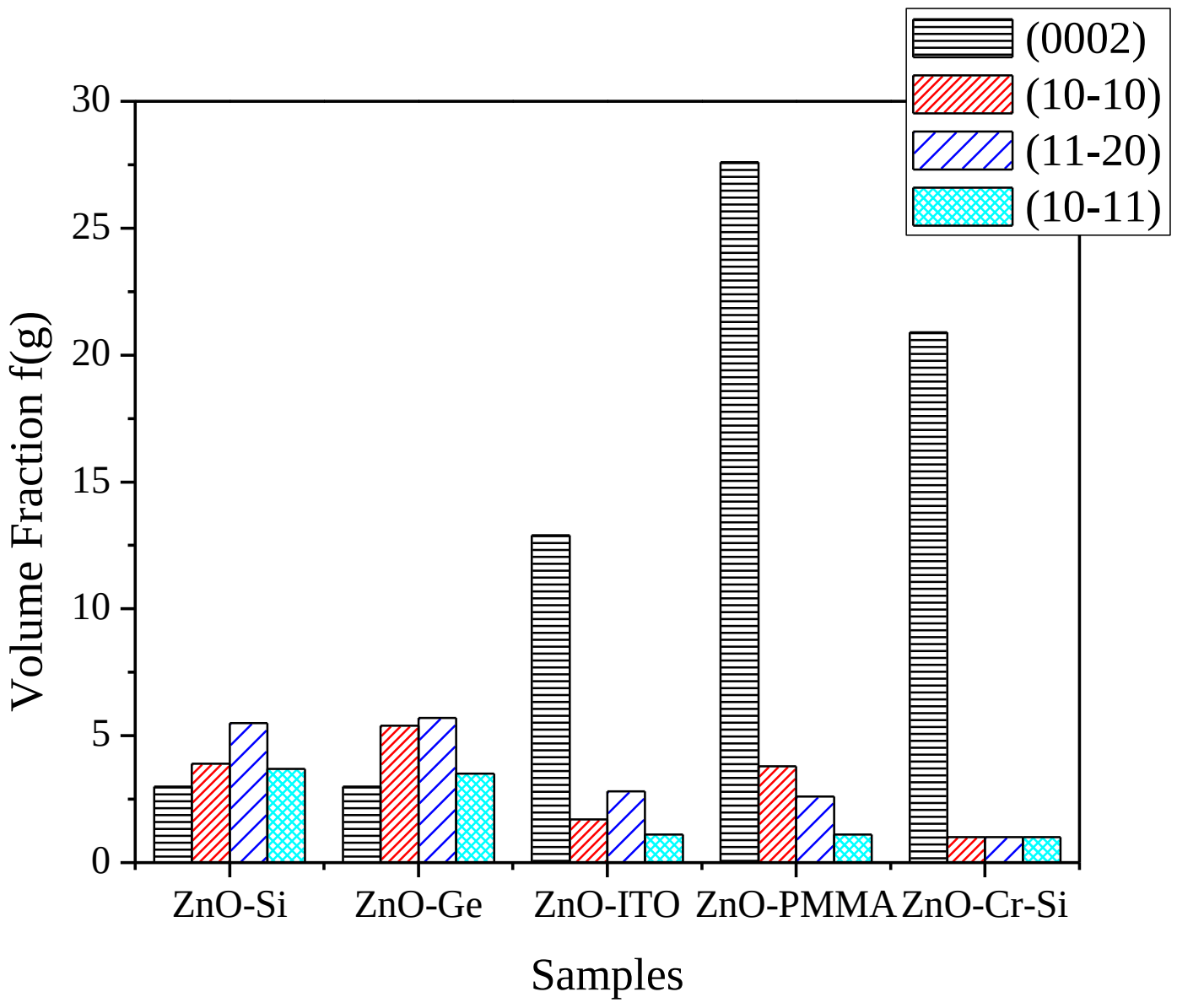

**Fig. 9.**